# Optimizing Deep Reinforcement Learning for American Put Option Hedging

Reilly Pickard[a]*, F. Wredenhagen[b], and Y. Lawryshyn[c],

[a]*Department of Mechanical and Industrial Engineering, University of Toronto, Toronto, Canada; [b]Ernst & Young LLP, Toronto, ON, M5H 0B3, Canada; [c]Department of Chemical Engineering, University of Toronto, Toronto, Canada;*

*Correspondence: reilly.pickard@mail.utoronto.ca

## Abstract

This paper contributes to the existing literature on hedging American options with Deep Reinforcement Learning (DRL). The study first investigates hyperparameter impact on hedging performance, considering learning rates, training episodes, neural network architectures, training steps, and transaction cost penalty functions. Results highlight the importance of avoiding certain combinations, such as high learning rates with a high number of training episodes or low learning rates with few training episodes and emphasize the significance of utilizing moderate values for optimal outcomes. Additionally, the paper warns against excessive training steps to prevent instability and demonstrates the superiority of a quadratic transaction cost penalty function over a linear version. This study then expands upon the work of Pickard et al. (2024), who utilize a Chebyshev interpolation option pricing method to train DRL agents with market calibrated stochastic volatility models. While the results of Pickard et al. (2024) showed that these DRL agents achieve satisfactory performance on empirical asset paths, this study introduces a novel approach where new agents at weekly intervals to newly calibrated stochastic volatility models. Results show DRL agents re-trained using weekly market data surpass the performance of those trained solely on the sale date. Furthermore, the paper demonstrates that both single-train and weekly-train DRL agents outperform the Black-Scholes Delta method at transaction costs of 1% and 3%. This practical relevance suggests that practitioners can leverage readily available market data to train DRL agents for effective hedging of options in their portfolios.

Keywords: reinforcement learning; neural networks; dynamic stock option hedging; derivatives pricing; quantitative finance; financial risk management

Subject classification code: 91G20

## 1. Introduction

In the sale of a put option, the seller faces the risk that the underlying asset price will drop, resulting in a payout to the buyer. As such, financial institutions seek a hedging strategy to offset the potential losses from a short put option position. A common option hedging strategy is the development of a Delta-neutral portfolio, which requires a position of Delta shares to be taken in the underlying, where Delta is the first partial derivative of the option price with respect to the underlying (Hull 2012). Delta hedging stems from the Black and Scholes (BS) (1973) option pricing model, which shows that a European option is perfectly replicated with a continuously rebalanced Delta hedge when the underlying asset price process follows a geometric Brownian motion (GBM) with constant volatility. However, financial markets operate in discrete fashion, volatility is ever-changing, and the impact of transaction costs need be considered. Further, many options are not European, such as American options in which there is a potential for early exercise.

Given the outlined market frictions, hedging an option position may be modelled as a sequential decision-making process under uncertainty. A method that has achieved success in such decision-making procedures is reinforcement learning (RL), a subfield of artificial intelligence (AI). Specifically, the combination of RL and neural networks (NNs), called deep RL (DRL), has been used to achieve super-human level performance in video games (Mnih *et al.* 2013), board games (Silver *et al.* 2014), and robot control (Lillicrap *et al*. 2015). Recent advances in quantitative finance have seen DRL be leveraged to achieve desirable results in hedging financial options, as described in the review provided Pickard and Lawryshyn (2023). Notably, prior work by (Pickard *et al.* 2024) showed the proficiency of DRL agents over the BS Delta strategy when hedging short American put options. Specifically, (Pickard *et al.* 2024) shows the following:

(1) When transaction costs are considered, DRL agents outperform the BS Delta and binomial tree hedge strategies when trained and tested on simulated paths from a GBM process

(2) When DRL agents are trained using paths from stochastic volatility models calibrated to market data, DRL agents outperform the BS Delta strategy on the realized asset path for the respective underlying.

While recent successes in the field of DRL option hedging are encouraging, particularly the results of Pickard *et al.* (2024), there is a lack of literature pertaining to best practices for financial institutions looking to implement this "black box" approach. For example, while many papers boast encouraging results, there is little discussion given to the selection of DRL model hyperparameters, which can have a large impact on DRL agent performance (Kiran and Ozyildirim 2022). In the field of option hedging, little discussion is given to hyperparameter choices. Of the 17 studies analysed in Pickard and Lawryshyn (2023), only Du et. al (2020), Assa et al. (2021), and Fathi & Hientzsch (2023) conduct some form of hyperparameter analysis. Moreover, the analysed studies from Pickard and Lawryshyn (2023) consider European style options, and it has been reported that slight environmental changes may impact the optimal hyperparameter settings (Henderson et al. (2017), Eimer et al. (2022)). As such, given that Pickard *et al.* (2024) provides a first DRL model dedicated to hedging American style options, this study will look to shed light on the hyperparameter tuning process, thereby optimizing the results in the process.

The first goal of this work is to provide general hyperparameter selection guidance for practitioners wishing to implement DRL hedging. Therefore, the results of this article will not only show what hyperparameter sets are optimal for the American option hedging task, but what combinations should be avoided. This study will first examine how learning rates, NN architectures, and the number of re-balance steps available to the agent in training effect DRL agent performance. Moreover, considering the results in Pickard *et al.* (2024), the first work to consider training agents for American options, this article will provide clarity on the key training choices such as the reward function to help the agent achieve optimal hedging. Finally, building off of the key result of Pickard *et al.* (2024), in that agents trained with market calibrated stochastic volatility model data outperform a BS Delta strategy on empirical asset paths, this article will examine the impact of re-training the agent at weekly intervals to newly available option data.

**2 Deep Reinforcement Learning**

RL, which is given a complete description by Sutton and Barto (2018), features an agent in some pre-defined environment. At each discrete time step $t$, the RL agent observes a state $S_t$, takes an action $A_t$ according to its policy, $\pi(S_t) \to A_t$, and receives a reward $R_{t+1}$. For a task with $T$ discrete steps, the object of the agent is to maximize the expected future rewards, $G_t = \sum_{k=0}^{T-1} \gamma^k R_{t+k+1}$, noting that $\gamma \in (0,1]$ is the discount factor. The RL goal is to learn the optimal policy, $\pi^*$, that maximizes the expected return. Policies are evaluated and improved via the $Q$-function:

$$Q^\pi(S_t, A_t) = \mathbb{E}_\pi[G_t | S_t, A_t] \in \mathbb{R}, \forall\, s \in \mathcal{S}, \forall\, a\, \mathcal{A}. \tag{1}$$

$\mathcal{S}$ and $\mathcal{A}$ are the state and action-spaces, respectively. RL problems iterate between policy evaluation, i.e., computing $Q^\pi$, and policy improvement. Policy improvement in tradition value-based RL methods transforms the policy as: $\pi(S_t) \to \arg\max Q^\pi(S_t, A_t)$. On-policy agents learn about the environment following this improved policy, which may lead to under-exploration, even if there is some probability that the next action is taken randomly. As such, Watkins (1989) devised an off-policy strategy in which a sub-optimal exploratory policy is used to generate experiences for updating $Q^\pi(S_t, A_t)$. Q-learning updates the current Q-value as:

$$Q^\pi(S_t, A_t) = Q^\pi(S_t, A_t) + \alpha \left( R_{t+1} + \gamma max_{A_{t+1}} Q^\pi(S_{t+1}, A_{t+1}) - Q^\pi(S_t, A_t) \right).$$

Updates of this style are intractable for high-dimensional problems, as the $Q$-function is tabular and requires the storage $\mathcal{S} \times \mathcal{A}$ values. However, a breakthrough from Mnih *et al.* (2013) shows that the $Q$-function may be approximated with a NN, recalling that this combination of NNs and RL is called DRL. Mnih *et al.* (2013) denote the $Q$-function as $Q(s, a; \theta)$, where $\theta$ is a vector of the NN parameters that signifies the weights and biases connecting the nodes within the neural network and use this method of deep Q-networks (DQN) to achieve super-human level performance in Atari games.

The NN is trained by optimized by minimizing the difference between the output and the target value. This objective function for iteration $i$ is given by

$$L_i(\theta_i) = \mathbb{E}_{s,a} \left[ (y_i - Q(S, A; \theta_i))^2 \right], \quad (2)$$

where $y_i = \mathbb{E}_{s,a} [R_{t+1} + \gamma max Q(S_{t+1}, A_{t+1}; \theta_{i-1})]$ is the target $Q$-value for iteration $i$ (Mnih *et al.* 2013). Based on this loss value, $\theta$ is updated via stochastic gradient descent (SGD):

$$\theta_i = \theta_{i-1} - \eta \nabla_{\theta_i} L_i(\theta_i). \quad (3)$$

Here $\eta$ is the learning rate for the NN, and $\nabla_{\theta_i} L_i(\theta_i)$ is the gradient of the objective function with respect to the network weights.

Although DQN is a breakthrough result for DRL, it is limited by its requirement for discrete action spaces, as the $Q$-function update require a max operation over all potential next actions. For a problem such as option hedging, a discrete action space limits the accuracy of potential hedging decisions. While hedging does require holding of a discrete number of shares, continuous action spaces are more prevalent in the DRL hedging literature, as the hedge is not limited to a discrete set of choices (Pickard and Lawryshyn 2023). Specifically, a prevalent algorithm in the DRL hedging space is deep deterministic policy gradient (DDPG), first detailed in Lillicrap *et al.* (2015). DDPG features two NNs, one for the critic and one for the actor. The critic network computes a $Q$-value for each action, and the network is trained by SGD (Lillicrap *et al.* 2015). While this is the same as DQN, the novelty of DDPG lies in the determination of the target, $y_i$, required for the computation of the objective function $L_i(\theta_i)$. In DDPG, the target is computed as $y_i = \mathbb{E}_{S,A}[R_{t+1} + \gamma Q(S_{t+1}, \pi_i(S_{t+1}); \theta_{i-1})]$ (Lillicrap *et al.* 2015), noting that the current policy estimate from the actor network, $\pi_i(S_{t+1})$ is used, rather than a max operation over $A_{t+1}$ as in DQN. This new update form eliminates the requirement of searching a discrete action space to update the $Q$-function. Further, by optimizing the actor network, an argmax operation is no longer required to improve the policy. Note that the actor network is optimized by gradient ascent, with the value of the starting state used as the objective function (Lillicrap *et al.* 2015).

**3 Similar Work**

*3.1 Option Hedging with Deep Reinforcement Learning*

The first to apply DRL in the option hedging space are Du *et al.* (2020) and Giurca and Borovkova (2021), who each use a DQN method. The literature is extended by Cao *et al.* (2021), who use the DDPG algorithm to allow for continuous action spaces. Assa *et al.* (2021), Xu and Dai (2022), Fathi and Hientzsch (2023), and Pickard *et al.* (2024). Cao *et al.* (2023) use distributed DDPG, which uses quantile regression to approximate the distribution of rewards. Across the literature, there a consistent inclusion of the current asset price, the time-to-maturity, and the current holding in the DRL agent state-space. However, there is a disagreement concerning the inclusion of the BS Delta in the state-space. Vittori *et al.* (2020), Giurca and Borovkova (2021), Xu and Dai (2022), Fathi and Hientzch (2023), and Zheng *et al.* (2023) include the BS Delta, while several articles argue that the BS Delta can be deduced from the option and stock prices and its inclusion only serves to unnecessarily augment the state (Cao *et al.* (2021), Kolm and Ritter (2019), Du *et al*. (2020)). Moreover, in Pickard *et al.* (2024), the DRL agent is attempting to hedge an American option, and therefore the inclusion of the BS Delta, which is derived for a European option, is counterproductive.

The mean-variance objective is a common reward function across the literature (Pickard and Lawryshyn 2023). A generalized mean-variance reward is given by $R_t = \delta w_t - \xi \delta w_t^2$, where $\xi$ is a measure of risk aversion, and $\delta w_t$ is the incremental wealth of the hedging portfolio at time $t$. Note the wealth increment is given by $\delta w_t = [C_t - C_{t-1}] + n_t[S_t - S_{t-1}] - c_t$, where $C_t$, is the option value, $n_t$ is the position in the underlying, and $c_t$ is a transaction cost penalty. Pickard *et al.* (2024) use a reward function that minimizes the difference between the option position value and the underlying position value, while also considering transaction costs.

The acquisition of underlying asset price data is an integral part of training a DRL option hedger, as the current price is required in both the state-space and the reward. Moreover, the option price is required for the reward function. In the DRL hedging literature, almost all studies employ at least one experiment wherein the data is generated through Monte Carlo (MC) simulations of a GBM process. As this is aligned with the BS model, studies that use GBM use the BS option price in the reward function (Pickard and Lawryshyn 2023). However, recall that Pickard *et al.* (2024) hedge American put options, and therefore compute the option price at each step by interpolating an American put option binomial tree. The majority of studies in the DRL option hedging literature also simulate training paths using a stochastic volatility model such as the SABR (Stochastic Alpha, Beta, Rho) (Hagan et al. 2002) or Heston (1993) models. Note that when the option is European, the option price may still be computed at each time with the BS model, simply by substituting the updated volatility into the closed form expression. However, in Pickard *et al.* (2024) the option is American and therefore the option price is computed by using a Chebyshev interpolation method. This Chebyshev method, which is given a through overview by Glau *et al.* (2018), is described in the next section. A final note on training data is that Pickard *et al.* (2024) are the only study to calibrate the stochastic volatility model to empirical option data. This calibration process is described in the next section, and the final experiment of this paper will examine the difference between one-time and weekly re-calibration to new market data.

Pickard and Lawryshyn (2023) describe that studies generally test the trained DRL agent using MC paths from the same underlying process from training, i.e., GBM or a stochastic volatility model. Consistent results are present in such experiments, as DRL agents who are trained with a transaction cost penalty outperform the BS Delta strategy when transaction costs are present (Pickard and Lawryshyn 2023). There are also cases wherein the DRL agents are tested on data that does not match the training data. For example, each of Giurca and Borovkova (2021), Xiao *et al.* (2021), Zheng *et al.* (2023), and Mikkilä and Kanniainen (2023) perform a sim-to-real test in which they test the simulation trained DRL agent on empirical data, and all but Giurca and Borovkova (2021) achieve desirable results.

Those using only empirical data for both training and testing are Pham *et al.* (2021) and Xu and Dai (2022). Pham *et al.* (2021) show that their DRL agent has a profit higher than the market return when testing with empirical prices. Of further note, Xu and Dai (2022) train and test their DRL agent with empirical American option data, which is not seen elsewhere in the DRL hedging literature. However, Xu and Dai (2022) do not change their DRL approach in any manner when considering American versus European options. As such, Pickard *et al.* (2024) are the first to train DRL agents to consider early exercise. Consistent with the literature, Pickard *et al.* (2024) first test agents with data generation processes that match the training, i.e., agents trained with MC paths of a GBM process are tested on MC paths of a GBM process, and agents trained with MC paths of the modified SABR model are tested on MC paths of the modified SABR model. Additionally, given that the modified SABR model is calibrated to market prices, a final test examines DRL agent performance on realized market paths. The results indicate that the DRL agents outperform the BS Delta method across 80 options. As this paper aims to optimize the results of Pickard *et al.* (2024), this process will be given a much more thorough description in the methodology section.

It is noted that while the literature featuring to the use of DRL for American option hedging is limited, there are still applications of DRL in American option applications. Fathan and Delage (2021) use double DQN to solve an optimal stopping problem for Bermudan put options, which they write are a proxy for American options. Ery and Michel (2021) apply DQN to solving an optimal stopping problem for an American option. While the DRL solutions are limited to the two studies described, further articles have used other RL techniques for American option applications such as pricing and exercising, and such articles are summarized in a review given by (Bloch 2023).

### *3.2 Hyperparameter Analysis*

As outlined above, only three of the 17 analysed DRL option hedging studies from Pickard and Lawryshyn (2023) detail some form of hyperparameter analysis. Du *et al.* (2020) conduct a grid search across multiple hyperparameter configurations, but only report the final hyperparameter set. The authors specifically report on their choice of discount factor, learning rates, and number of hidden layers. Assa *et al.* (2021) also perform a hyperparameter sweep, but more insight is provided into how poor choices can negatively affect performance. They first test a range of training episode lengths, varying from 12k to 45k, and report that after 20k training episodes, performance begins to decline and eventually diverges. Further, Assa *et al.* (2021), who employ DDPG and compare results across actor and critic network sizes of $16^3$, $64^3$, $64^4$, and $64^5$, and find that no improvement occurs beyond a size of $64^4$. Fathi and Hientzsch (2023) conduct a sensitivity analysis on the choice of discount factor, learning rates, and NN architectures. The authors, using DDPG, find minimal differences across discount factors, and conclude that to 1e-6 and 1e-5 is the optimal actor-critic learning rate combination for their environment. Further, Fathi and Hientzch (2023) detail that best results come from the deepest NN architecture (5 layers).

This lack of discussion given to hyperparameters hinders the integration of DRL into real world applications, as results are difficult to reproduce, specifically in complex financial environments. While the financial field has little literature focused on DRL hyperparameters, several studies have been conducted to assess the impact of training decisions for general DRL tasks. Henderson *et al.* (2017) use the HalfCheetah, Hopper, and Swimmer environments from the OpenAI gym module developed by (Brockman *et al.* 2016*)*. The authors conduct experiments with DDPG, proximal policy optimization (PPO), and trust region policy optimization (TRPO) algorithms, and investigate the effects of network architecture, reward scaling, and random seed initialization across the different gym environments. For the DDPG case, they test NN architectures of $64^2$, 100×50×25, and 50×25, and find that the effects are not consistent across the gym environments. Moreover, the authors find that even if all other hyperparameters are held constant, modifying the random seed initialization for the training process can drastically impact results.

Islam *et al.* (2017) follow up on the work of Henderson *et. al* (2017) by examining the performance of DDPG and TRPO methods in the HalfCheetah and Hopper gym environments. They first note that it is difficult to reproduce the results of Henderson *et al.* (2017), even with similar hyperparameter configurations. Next, Islam *et al.* (2017) add to the literature an analysis of DDPG actor and critic learning rates. They conclude first that the optimal learning rates vary between the Hopper and HalfCheetah environments, before noting it is difficult to gain a true understanding of optimal learning rate choices while all other parameters are held fixed. Andrychowicz *et al.* (2020) perform a thorough hyperparameter sensitivity analysis for multiple on-policy DRL methods, but do not consider off-policy methods such as DDPG. Overall, Andrychowicz *et al.* (2020) conclude that performance is highly dependent on hyperparameter tuning, and this limits the pace of research advances. Several other studies find similar results when testing various DRL methods in different environments, whether through a manual search of hyperparameters or some pre-defined hyperparameter optimization algorithm (Ashraf *et al.* (2021), Kiran and Ozyildirim (2022), Eimer *et al.* (2022)).

As such, it is evident that hyperparameter configurations can drastically impact DRL agent results. However, much of the literature on DRL hyperparameter choices often utilize a generic, pre-constructed environment for analysis, rather than addressing real-life applicable problems. Therefore, this study aims to contribute to the DRL hedging literature, and the DRL space as a whole, by conducting a thorough investigation of how hyperparameter choices impact the realistic problem of option hedging in a highly uncertain financial landscape.

# 4 Methodology

## 4.1 General DRL Agent Setup

All experiments in this study employ the DDPG algorithm to train DRL agents. Moreover, the rectified linear unit (ReLu) activation function is used as the activation function for both the actor and critic hidden layers. The critic output is linear, and the actor output is transformed by a sigmoid and multiplied by $-1$ to map the output to the range $[0, -1]$, which aligns with the desired hedging action for a short American put. The DRL state-space is the same as Pickard *et al.* (2024) and consists of the current asset price, the time-to-maturity, and the current holding (previous action). The rest of the hyperparameters, i.e., the actor and critic learning rates, the actor and critic NN architectures, the number of episodes, and the steps per episode, will vary across the experiments. The reward function will also vary across experiments, with different transaction cost penalties being investigated. These hyperparameter analyses are detailed in the next subsection.

## 4.2 Hyperparameter Experiments

The first round of analysis in this article is focused on examining hyperparameter impact on DRL American option hedging performance. For all hyperparameter experiments, training data is in the form of simulated paths of the GBM process. The time-to-maturity is one year, the option is at the money with a strike of $100, the volatility is 20%, and the risk-free-rate is equal to the mean expected return of 5%. All tests also use GBM data, and each test consists of 10k episodes, consistent with Pickard *et al.* (2024). Note that to assess the robustness and consistency of DRL agents, testing will be performed with transaction cost rates of both 1 and 3%. Table 1 summarizes the hyperparameters used in Pickard *et al.* (2024), and for each experiment, the

hyperparameters not being analysed are held fixed at these values. This set of hyperparameters is referred to as the base case.

Table 1: Base Case Hyperparameter Summary

| | |
|---|---|
| Actor Learning Rate | 5e-6 |
| Critic Learning Rate | 5e-4 |
| Training Episodes | 5000 |
| Steps per Training Episode | 25 |
| Actor NN Architecture | $64^2$ |
| Critic NN Architecture | $64^2$ |
| TC Penalty Function | $\kappa(A_t - A_{t-1})^2 S_t$ |
| TC Penalty Multiplier ($\kappa$) | 0.005 |

A first experiment will examine how DRL agent performance is impacted by actor and critic learning rates as a function of episodes. Specifically, a grid search is performed using actor learning rates of 1e-6, 5e-6, and 10e-6, critic learning rates of 1e-4, 5e-4, and 10e-4, and episode lengths of 2500, 5000, and 7500. A second experiment will assess the impact of neural-network architectures. Similar to the first experiment, tests will be conducted across actor learning rates of 1e-6, 5e-6, and 10e-6, critic learning rates of 1e-4, 5e-4, and 10e-4, and NN architectures of $32^2$, $64^2$, and $64^3$, for both the actor and critic networks, respectively.

Next, an experiment is conducted to evaluate the impact of training steps. Here, DRL agents are trained using 10, 25, and 50 steps over the year of hedging. Moreover, to examine whether there is a performance impact based on the difference between training steps and testing re-balance periods, testing will be conducted by considering environments with 52 (weekly), 104 (bi-weekly), and 252 (daily) re-balancing times. To be clear, in this study, the steps per episode is synonymous with re-balance periods. As such, for this particular hyperparameter experiment, the training is comprised of 10, 25, and 50 rebalance periods (steps), spaced equally throughout the episode (which has length of 1 year), and the testing is comprised of 52, 104, and 252 rebalance periods (steps).

In the final hyperparameter experiment, the effect of the transaction cost penalty function in the reward is assessed. While much of the European option hedging literature uses a linear function, $\psi|A_t - A_{t-1}|S_t$, where $\psi$ is the linear transaction cost penalty multiplier, Pickard *et al.* (2024) employ a quadratic penalty, $\kappa(A_t - A_{t-1})^2 S_t$, for hedging American put options. To examine the difference, agents will first be trained and tested using the linear function for $\psi$ = [0.001, 0.05, 0.01, 0.03], and these agents will be compared to those trained with the quadratic function for $\kappa$ = [0.001, 0.05, 0.01], but still tested using the more realistic linear function. Note that the $\psi$ = 0.03 is added to the linear experiments as this aligns exactly with the testing transaction cost function, $\lambda|A_t - A_{t-1}|S_t$, for the case where $\lambda$ = 3%.

## *4.3 Optimization of Market Calibrated DRL Agents*

After the hyperparameter analysis, this study looks to improve on the key result from Pickard *et al.* (2024), wherein it is shown that DRL agents trained with paths from a market calibrated stochastic volatility model outperform DRL agents on the true asset paths between the option sale and maturity dates. One key shortcoming in the findings from Pickard *et al.* (2024) is that the stochastic volatility model is only calibrated to the option data on the sale date, and the agent trained with model is used to hedge the entire life of the option. As such, if any cases presented themselves where the market changed drastically between the sale and the maturity, this would render the agent obsolete, as it had been trained to hedge in a completely different environment. As such, it is hypothesized that performance will improve by re-training agents to new data at weekly intervals.

Given that the underlying asset data now stems from a stochastic volatility model, an alternative method is required for providing the DRL agent reward with the option price at each training step. While in the GBM experiments, the option price required for the DRL agent reward on each training step is obtained by interpolating a binomial tree, tree methods for stochastic volatility processes increase in complexity due to the requirement of a second spatial dimension. As such, Pickard *et al.* (2024) detail a Chebyshev interpolation method, first introduced in Glau *et al.* (2018), to provide the agent with the American option price at each time step in training.

Chebyshev interpolation involves weighting orthogonal interpolating polynomials to approximate a known function within predetermined upper and lower bounds at each time step. Computing an American option price via Chebyshev interpolation involves a three-step process:

1. Discretize the computational space from $t_0 = 0$ to $t_N = T$. Then, Chebyshev nodes are generated between the upper and lower bounds at $t_{N-1}$ and paths are simulated from these nodes to maturity at $t_N$. The continuation value is computed as the discounted average payoff for all nodes at $t_{N-1}$ The value function for Chebyshev nodes at $t_{N-1}$ is determined by comparing the continuation value with the immediate exercise payoff.

2. For the next time step, $t_{N-2}$, the process is repeated by simulating paths from $t_{N-2}$ to $t_{N-1}$ and performing Chebyshev interpolation between nodes at $t_{N-1}$. This continues backward in time until $t = t_0$, and by indexing optimal exercise nodes, a Chebyshev exercise boundary is computed.

3. After completing steps 1 and 2, value functions become accessible at every node throughout the discretized computational space. Consequently, with a specified asset price level and time step, Chebyshev interpolation can be utilized to calculate the price of an option. This step may be called at any time step in the DRL agent training process, without repeating steps 1 and 2.

It is noted first that the Chebyshev approach is agnostic to the underlying asset evolution process, allowing for easy transitions to more complex environments.

A common alternative for computing option prices for a stochastic volatility case is to simulate several thousand MC paths from the current asset price level to maturity or exercise, where the exercise boundary is often computed using the Longstaff-Schwartz Monte Carlo (LSMC) method (Longstaff and Schwartz 2001). However, incorporating a simulation-based pricing method into the training of a reinforcement learning (RL) agent would substantially prolong the training duration, as each training step would necessitate a new set of simulations. For instance, in this study, a base case training loop comprises 5000 episodes, each spanning 25 steps. Consequently, a total of 125,000 sets of simulations would be requiring, an unnecessary addition to an already time-intensive procedure of training a neural network. For a more complete description of the Chebyshev method and its advantages over simulation-based LSMC pricing, the reader is directed to Glau *et al.* (2018).

As described, this studies employs a strategy wherein each week, a new model is calibrated, new Chebyshev nodes are generated for pricing, and a DRL agent is trained to hedge according to the current market conditions. Note that the stochastic volatility model is the same as in Pickard *et al.* (2024), which is an adaptation of the SABR model (Hagan *et al.* 2002):

$$\sigma_t = \sigma_{t-1} + \nu \sigma_{t-1} \Delta B_t,$$
$$S_t = S_{t-1} + \mu S_{t-1} \Delta t + \sigma_t S_{t-1} \Delta W_t,$$
(4)

where $B_t$ and $W_t$ are Brownian motions with increments given by

$$\begin{bmatrix} \Delta W_t \\ \Delta B_t \end{bmatrix} = \sqrt{\Delta t} \begin{bmatrix} Z_1 \\ \rho Z_1 + \sqrt{1-\rho^2} Z_2 \end{bmatrix}, \tag{5}$$

noting that $Z_1$ and $Z_2$ are independent standard normal random variables. In this stochastic volatility model, $\rho$ and $\nu$ are model coefficients that may be used to calibrate the model to empirical observations. Specifically, option data is retrieved for 8 symbols and 5 strikes each. The first training date is October 16th, 2023, and the maturity for all options is November 17th, 2023. In one round of training, DRL agents are trained only using paths from the stochastic volatility model calibrated to data on October 16th. In a second round, DRL agents are trained to hedge only a week at a time. One agent is trained using the data from Monday, October 16th and to hedge only 5 days into the future. A new agent is trained from data available on Monday, October 23rd, and is again trained to hedge for 5 days. The process repeats, with new agents trained using data from October 30th, November 6th, and November 13th, respectively.

A comparison is then given by testing agents on the true asset paths between October 16th and November 17th, using 1 and 3% transaction costs. Note that the BS Delta method is used as a benchmark. Specifically, the BS delta uses the volatility from the weekly re-calibrated stochastic volatility model. Table 2 summarizes the option data. The base case hyperparameters are used for this analysis.

Table 2: Option Data Summary

| Symbol | Strikes |
|---|---|
| GE | $100, $105, $110, $115, $120 |
| XOM | $100, $105, $110, $115, $120 |
| DELL | $60, $65, $70, $75, $80 |
| PEP | $150, $155, $160, $165, $170 |
| AMZN | $125, $130, $135, $140, $145 |
| TSLA | $210 $215, $220, $225, $230 |
| TXN | $135, $140, $145, $150, $155 |
| AIG | $50, $55, $60, $65, $70 |

# 5 Results

## 5.1 Hyperparameter Analysis

The first experiment analyses the impact of actor learning rates, critic learning rates, and training episodes. Table 3 summarizes the mean and standard deviation of hedging P&L for 1 and 3% transaction cost rates. For comparison, note that the table also lists the BS Delta performance. The binomial tree hedge is not listed as it is worse than the BS Delta under both 1 and 3% transaction costs. Note that the italicized and underlined results are for agents that outperform the BS Delta in both mean and standard deviation, under both 1 and 3% transaction costs.

Table 3: Final P&L Statistics across Learning Rate – Training Episode Combinations.

| $\lambda = 1\%$ (BS Delta: Mean = -$3.13, SD = $1.22) | | | | | | | |
|---|---|---|---|---|---|---|---|
| **Training Episodes** | | **2500** | | **5000** | | **7500** | |
| **Actor Learning Rate** | **Critic Learning Rate** | **Mean** | **SD** | **Mean** | **SD** | **Mean** | **SD** |
| 1e-6 | 1e-4 | -$0.63 | $5.74 | -$3.16 | $1.56 | -$2.93 | $1.42 |
| 1e-6 | 5e-4 | -$1.23 | $3.28 | -$2.21 | $1.90 | -$2.89 | $2.59 |
| 1e-6 | 10e-4 | -$1.02 | $3.63 | -$2.65 | $1.70 | -$2.89 | $1.41 |
| 5e-6 | 1e-4 | -$6.10 | $6.98 | -$2.91 | $1.60 | -$3.15 | $1.56 |
| 5e-6 | 5e-4 | -$3.12 | $1.61 | _-$2.65_ | _$1.12_ | -$2.97 | $1.33 |
| 5e-6 | 10e-4 | -$2.70 | $1.31 | -$3.40 | $1.81 | -$3.16 | $3.12 |
| 10e-6 | 1e-4 | -$3.54 | $2.16 | -$3.40 | $2.15 | -$2.26 | $1.65 |
| 10e-6 | 5e-4 | -$2.99 | $1.25 | -$2.74 | $1.54 | -$3.21 | $1.99 |
| 10e-6 | 10e-4 | -$2.99 | $1.64 | -$3.28 | $2.26 | -$3.59 | $3.99 |
| $\lambda = 3\%$ (BS Delta: Mean = -$8.86, SD = $3.21) | | | | | | | |
| 1e-6 | 1e-4 | -$1.14 | $5.49 | -$9.05 | $3.03 | -$8.36 | $2.90 |
| 1e-6 | 5e-4 | -$3.18 | $2.89 | -$6.26 | $2.48 | -$8.34 | $4.54 |
| 1e-6 | 10e-4 | -$2.72 | $3.45 | -$7.51 | $2.65 | -$8.14 | $2.76 |
| 5e-6 | 1e-4 | -$17.50 | $14.74 | -$8.20 | $3.05 | -$9.08 | $3.69 |
| 5e-6 | 5e-4 | -$8.99 | $3.65 | _-$7.51_ | _$2.46_ | -$8.49 | $3.38 |
| 5e-6 | 10e-4 | -$7.70 | $2.82 | -$9.90 | $4.87 | -$8.92 | $5.15 |
| 10e-6 | 1e-4 | -$10.17 | $4.84 | -$9.66 | $5.00 | -$6.41 | $1.74 |
| 10e-6 | 5e-4 | -$8.59 | $2.90 | -$7.77 | $2.73 | -$9.08 | $3.71 |
| 10e-6 | 10e-4 | -$8.45 | $3.47 | -$9.47 | $4.40 | -$10.39 | $7.26 |

The results show that the actor learning rate – critic learning rate – episode combination of (5e-6, 5e-4, 5000) is optimal, outperforming the BS Delta in terms of mean and standard deviation of final P&L under both 1 and 3% transaction costs. The results of Table 3 also gives other key insights into the impact of learning rates and episodes. Take for example the case of low learning rates and low episodes, i.e., combinations of (1e-6, 1e-4, 2500) and (1e-6, 5e-4, 2500). Table 3 shows that these agents perform well in terms of achieving high mean final P&L's, but standard deviation results are not desirable. This indicates that given the low learning rates and episodes, the agents do not learn a consistent hedging strategy that yields the desired low standard deviation. Now take for example cases wherein learning rates and episodes are at the high end, i.e., a combination of (10e-6, 5e-4, 7500). These agents do not outperform the BS Delta mean or standard deviations in any cases. As such, it is likely that due to the high rate of learning and higher number of training episodes, the training environment becomes unstable. These results illustrate that keeping the learning rates and episodes in a middling range such as (5e-6, 5e-4, 5000) is yields satisfactory performance, beating Delta in mean and standard deviation at both 1 and 3% transaction costs. However, when the critic learning rate is kept low (i.e, 1e-4), increasing the number of episodes often leads to improved performance, as one would expect. This is observed by the combination of (10e-6, 1e-4, 7500), which is considerably better than the (5e-6, 5e-4, 5000) at 3% transaction costs, and has a lower mean at 1% transaction costs. These results suggest that if one can find an optimal learning rate combination for the given problem, increasing the number of episodes will further the learning of the DRL agent and an improved policy will result.

Table 4 details the results of the next experiment, wherein the learning rates and NN architectures for both the actor and critic networks are varied. Note again that italicized and underlined results are for agents that outperform the BS Delta in both mean and standard deviation, under both 1 and 3% transaction costs. Further, the reader should be aware that row and column IDs are added to Table 4 due to the amount of data presented, to aid in the explanation of results.

Table 4: Final P&L Statistics across Learning Rate – NN Architecture Combinations

| | Column ID | | A | B | C | D | E | F |
|---|---|---|---|---|---|---|---|---|
| | $\lambda = 1\%$ (BS Delta: Mean = -$3.13, SD = $1.22) | | | | | | | |
| | | Critic Architecture | $32^2$ | | $64^2$ | | $64^3$ | |
| Row ID | Actor Architecture | Actor Learning Rate | Critic Learning Rate | Mean | SD | Mean | SD | Mean | SD |
| | | | | Mean | SD | Mean | SD | Mean | SD |
| 1 | | 1e-6 | 1e-4 | -$1.03 | $3.48 | -$1.50 | $2.99 | -$1.86 | $2.00 |
| 2 | | 1e-6 | 5e-4 | -$0.69 | $4.09 | -$1.32 | $2.89 | -$1.52 | $2.75 |
| 3 | | 1e-6 | 10e-4 | -$1.07 | $3.52 | -$1.62 | $2.67 | -$1.64 | $2.91 |
| 4 | | 5e-6 | 1e-4 | -$2.06 | $1.85 | -$2.75 | $1.94 | -$2.70 | $1.52 |
| 5 | $32^2$ | 5e-6 | 5e-4 | -$2.86 | $1.32 | -$3.26 | $2.01 | -$2.89 | $1.84 |
| 6 | | 5e-6 | 10e-4 | -$3.13 | $1.36 | -$3.04 | $1.56 | -$2.96 | $1.40 |
| 7 | | 10e-6 | 1e-4 | -$3.25 | $1.53 | -$3.28 | $2.04 | -$3.38 | $1.71 |
| 8 | | 10e-6 | 5e-4 | -$5.75 | $6.94 | -$3.41 | $1.63 | -$2.96 | $1.65 |
| 9 | | 10e-6 | 10e-4 | -$3.37 | $2.36 | -$3.22 | $1.73 | -$2.55 | $2.11 |
| 10 | | 1e-6 | 1e-4 | -$0.42 | $7.08 | -$3.16 | $1.56 | -$2.77 | $1.48 |
| 11 | | 1e-6 | 5e-4 | -$2.77 | $1.57 | -$2.21 | $1.90 | -$2.76 | $1.76 |
| 12 | | 1e-6 | 10e-4 | -$3.13 | $1.77 | -$2.65 | $1.70 | -$2.93 | $1.95 |
| 13 | | 5e-6 | 1e-4 | -$3.23 | $1.70 | -$2.91 | $1.60 | -$3.23 | $1.74 |
| 14 | $64^2$ | 5e-6 | 5e-4 | -$3.29 | $2.56 | *-$2.65* | *$1.12* | -$3.02 | $1.54 |
| 15 | | 5e-6 | 10e-4 | -$2.94 | $1.30 | -$3.40 | $1.81 | -$2.87 | $1.36 |
| 16 | | 10e-6 | 1e-4 | -$2.23 | $1.97 | -$3.40 | $2.15 | *-$2.97* | *$1.14* |
| 17 | | 10e-6 | 5e-4 | -$3.63 | $2.08 | -$2.74 | $1.54 | -$2.74 | $1.87 |
| 18 | | 10e-6 | 10e-4 | -$3.31 | $1.80 | -$3.28 | $2.26 | -$3.31 | $3.39 |
| 19 | | 1e-6 | 1e-4 | -$2.57 | $1.46 | -$3.01 | $1.80 | -$2.83 | $2.61 |
| 20 | $64^3$ | 1e-6 | 5e-4 | -$3.02 | $1.32 | -$2.58 | $2.44 | -$2.91 | $1.79 |
| 21 | | 1e-6 | 10e-4 | -$2.78 | $2.00 | -$2.27 | $2.29 | -$3.21 | $2.44 |

| | | | | | | | | |
|---|---|---|---|---|---|---|---|---|
| 22 | | 5e-6 | 1e-4 | -$2.30 | $1.40 | -$3.40 | $1.81 | -$3.39 | $1.82 |
| 23 | | 5e-6 | 5e-4 | -$2.08 | $1.85 | -$3.44 | $1.65 | -$2.79 | $2.09 |
| 24 | | 5e-6 | 10e-4 | -$3.14 | $1.66 | -$3.60 | $2.52 | -$2.97 | $2.04 |
| 25 | | 10e-6 | 1e-4 | -$3.61 | $2.24 | -$3.22 | $1.64 | -$3.55 | $1.99 |
| 26 | | 10e-6 | 5e-4 | -$2.71 | $1.75 | -$3.89 | $2.51 | -$3.15 | $1.57 |
| 27 | | 10e-6 | 10e-4 | -$3.88 | $2.26 | -$4.00 | $2.90 | -$3.31 | $3.50 |
| | $\lambda = 3\%$ (BS Delta: Mean = -$8.86, SD = $3.21) | | | | | | | | |
| 28 | | 1e-6 | 1e-4 | -$2.53 | $3.19 | -$4.07 | $2.49 | -$5.20 | $1.83 |
| 29 | | 1e-6 | 5e-4 | -$1.78 | $3.87 | -$3.47 | $2.52 | -$4.11 | $2.53 |
| 30 | | 1e-6 | 10e-4 | -$2.99 | $3.27 | -$4.49 | $2.33 | -$4.35 | $2.39 |
| 31 | | 5e-6 | 1e-4 | -$5.65 | $1.61 | -$7.77 | $3.33 | -$7.64 | $2.81 |
| 32 | $32^2$ | 5e-6 | 5e-4 | -$8.17 | $2.59 | -$9.44 | $3.93 | -$8.17 | $3.34 |
| 33 | | 5e-6 | 10e-4 | -$9.00 | $3.38 | -$8.61 | $3.47 | -$8.43 | $3.39 |
| 34 | | 10e-6 | 1e-4 | -$9.35 | $3.57 | -$9.33 | $4.45 | -$9.60 | $4.08 |
| 35 | | 10e-6 | 5e-4 | -$16.58 | $14.58 | -$9.66 | $3.88 | -$8.36 | $3.54 |
| 36 | | 10e-6 | 10e-4 | -$9.69 | $5.05 | -$9.07 | $4.26 | -$7.30 | $2.77 |
| 37 | | 1e-6 | 1e-4 | -$0.87 | $7.03 | -$9.05 | $3.03 | -$7.89 | $2.92 |
| 38 | | 1e-6 | 5e-4 | -$7.86 | $2.34 | -$6.26 | $2.48 | -$7.82 | $3.05 |
| 39 | | 1e-6 | 10e-4 | -$8.91 | $3.78 | -$7.51 | $2.65 | -$8.37 | $3.80 |
| 40 | | 5e-6 | 1e-4 | -$9.26 | $3.49 | -$8.20 | $3.05 | -$9.27 | $4.24 |
| 41 | $64^2$ | 5e-6 | 5e-4 | -$9.54 | $4.86 | *-$7.51* | *$2.46* | -$8.64 | $3.70 |
| 42 | | 5e-6 | 10e-4 | -$8.34 | $2.70 | -$9.90 | $4.87 | -$8.24 | $3.55 |
| 43 | | 10e-6 | 1e-4 | -$6.32 | $1.84 | -$9.66 | $5.00 | *-$8.48* | *$2.92* |
| 44 | | 10e-6 | 5e-4 | -$10.43 | $4.88 | -$7.77 | $2.73 | -$7.76 | $3.21 |
| 45 | | 10e-6 | 10e-4 | -$9.55 | $4.17 | -$9.47 | $4.40 | -$9.40 | $5.93 |
| 46 | | 1e-6 | 1e-4 | -$7.25 | $2.85 | -$8.50 | $3.73 | -$8.02 | $4.21 |
| 47 | | 1e-6 | 5e-4 | -$8.56 | $3.26 | -$7.27 | $3.10 | -$8.31 | $3.67 |
| 48 | | 1e-6 | 10e-4 | -$7.84 | $2.45 | -$6.40 | $2.32 | -$9.08 | $4.01 |
| 49 | | 5e-6 | 1e-4 | -$6.38 | $2.11 | -$9.70 | $3.91 | -$9.63 | $3.68 |
| 50 | $64^3$ | 5e-6 | 5e-4 | -$5.81 | $2.63 | -$9.89 | $4.11 | -$8.05 | $3.76 |
| 51 | | 5e-6 | 10e-4 | -$8.99 | $3.61 | -$10.37 | $5.71 | -$8.54 | $4.68 |
| 52 | | 10e-6 | 1e-4 | -$10.35 | $5.31 | -$9.18 | $3.12 | -$10.21 | $4.94 |
| 53 | | 10e-6 | 5e-4 | -$7.78 | $2.24 | -$11.27 | $5.30 | -$8.97 | $3.13 |
| 54 | | 10e-6 | 10e-4 | -$11.19 | $6.02 | -$11.46 | $6.94 | -$9.54 | $5.69 |

Just as in the previous experiment, Table 4 gives insight on which combinations yield poor performance and should be avoided. Specifically, the results first show that actor and critic architecture of size $32^2$ are generally too shallow for this problem. When both the actor and critic NNs have a $32^2$ (rows 1-9, 28-36, and columns A-B) the results are at their worst, rarely outperforming the BS delta in mean or standard deviation, under both 1 and 3% transaction costs. However, while the shallowest combination of actor and critic NNs yields the worst results, the deepest combination ($64^3$ and $64^3$) does not necessarily yield the best results for this problem. Specifically, poor results are observed when using a deep actor NN with a high actor learning rate (rows 25-27, 52-54). This generally indicates the deep architecture in conjunction with large weight updates at each iteration yields instability, similar to the case wherein high learning rates were used in conjunction with high training episodes.

As for positive results, recall that in the previous experiment, (5e-6, 5e-4, 5000) was the only combination that achieved better results than the BS Delta on both 1 and 3 % transaction costs. However, in the previous experiment the actor-critic learning rate combination of 10e-6 and 1e-4 also yielded satisfactory results when using 7500 episodes. Now, when varying the NN architectures, it is shown that when actor critic learning rates are 10e-6 and 1e-4, respectively, increasing the critic NN architecture to $64^3$ yields a DRL agent that outperforms the BS Delta on both 1 and 3% transaction costs. As such, just as it was seen that lowering the critic learning rate and increasing the episodes yields positive performance, lowering the critic learning rate and increasing the depth of the NN improves hedging results. These results show that if the learning rates are optimized, i.e., lowered sufficiently, expanding the complexity of training, whether via increased episodes or a deeper NN will improve performance. However, if the learning rates are too high, increasing the model complexity will often yield worse results.

Table 5 displays the results of the next test, wherein DRL agents are tested using 10, 25, or 50 steps, and test in environments with 52 (weekly), 104 (bi-weekly), and 252

(daily) re-balancing steps. Final P&L statistics for the BS Delta benchmark are also listed for each number of re-balancing steps.

Table 5: Final P&L Statistics across Training Step – Testing Step Combinations

| | \multicolumn{6}{c}{$\lambda = 1\%$} | | | | | |
|---|---|---|---|---|---|---|
| **Testing Steps:** | 52 | | 104 | | 252 | |
| | **Mean** | **SD** | **Mean** | **SD** | **Mean** | **SD** |
| **BS Delta** | -$2.34 | $1.21 | -$3.16 | $1.21 | -$4.71 | $1.71 |
| **10 Training Steps** | -$1.81 | $1.78 | -$2.52 | $1.54 | -$3.84 | $1.33 |
| **25 Training Steps** | -$1.92 | $1.26 | -$2.70 | $1.12 | -$4.08 | $1.29 |
| **50 Training Steps** | -$2.25 | $3.66 | -$3.19 | $3.92 | -$4.80 | $4.57 |
| | \multicolumn{6}{c}{$\lambda = 3\%$} | | | | | |
| **BS Delta** | -$6.10 | $2.36 | -$8.59 | $3.37 | -$13.14 | $5.49 |
| **10 Training Steps** | -$5.04 | $1.80 | -$7.14 | $1.91 | -$11.06 | $2.82 |
| **25 Training Steps** | -$5.42 | $1.94 | -$7.61 | $2.55 | -$11.77 | $4.01 |
| **50 Training Steps** | -$6.32 | $5.15 | -$9.00 | $6.39 | -$13.63 | $8.66 |

Immediately evident from the results is that for the actor learning rate – critic learning rate – episode combination held fixed for this experiment, (5e-6, 5e-4, 5000), 50 training steps yields an unstable training process, as the training data is increased without lowering the learning rate. In all cases, the mean and standard deviation for the agent trained using 50 steps is the worst among the three DRL agents. As for the other two agents, trained with 10 and 25 steps, respectively, it seems as if their results are quite similar. In all cases, the agent trained with 10 steps has a higher mean final P&L, but the agent trained with 25 steps always has a lower standard deviation. This indicates that more than 10 training steps are required to achieve a robust agent with consistent hedging performance with low variability. As for the effect of testing re-balance periods, there seems to be no general performance relationship between training and testing steps. For example, the agent trained with 10 steps is still able to perform quite well when there are 252 re-balance periods in testing. This is an encouraging result, as practitioners looking to implement agents to hedge frequently do not need to significantly increase the training steps, which would yield longer training times. Another note is that as re-balancing frequency increases, the discrepancy between DRL and Delta performance is more pronounced. This is because of transaction costs: more re-balancing yields more transaction cost accumulation, and it has been shown repetitively throughout the literature that DRL agents outperform the BS Delta as transaction cost effects become more pronounced.

The final test for the hyperparameter analysis is an investigation of how the reward function transaction penalty impacts results. Specifically, we compare agents trained with a linear penalty, $\psi|A_t - A_{t-1}|S_t$, for $\psi = [0.001, 0.05, 0.01, 0.03]$, and those trained with a quadratic penalty, $\kappa(A_t - A_{t-1})^2 S_t$, for $\kappa = [0.001, 0.05, 0.01]$. Table 6 summarizes the results. Note that italicized and underlined values are for agents that outperform the BS Delta in terms of both a higher mean final P&L and a lower standard deviation.

Table 6: Final P&L Statistics across Varied Transaction Cost Penalty Functions

|  | $\lambda = 1\%$ | | $\lambda = 3\%$ | |
|---|---|---|---|---|
|  | **Mean** | **SD** | **Mean** | **SD** |
| **BS Delta** | -$3.13 | $1.22 | -$8.86 | $3.21 |
| **Linear TC Penalty: $\psi|A_t - A_{t-1}|S_t$** | | | | |
| $\psi = 0.001$ | -$3.30 | $2.41 | -$9.67 | $4.76 |
| $\psi = 0.005$ | -$3.14 | $1.64 | -$9.02 | $3.99 |
| $\psi = 0.01$ | -$3.12 | $1.87 | -$8.90 | $4.29 |
| $\psi = 0.03$ | -$11.22 | $13.23 | -$33.43 | $33.07 |
| **Quadratic TC Penalty: $\kappa(A_t - A_{t-1})^2 S_t$** | | | | |
| $\kappa = 0.001$ | -$3.42 | $2.47 | -$9.77 | $4.94 |
| $\kappa = 0.005$ | _-$2.65_ | _$1.12_ | _-$7.51_ | _$2.46_ |
| $\kappa = 0.01$ | _-$2.67_ | _$1.10_ | _-$7.56_ | _$2.27_ |

The results confirm the hypothesis from Pickard *et al.* (2024) in that DRL agents trained with a quadratic transaction cost penalty achieve superior performance in all cases. To reiterate, a quadratic penalty was implemented to punish large changes in the hedging position more heavily. This was done due to the threat of early exercise, and the quadratic penalty still enables the agent to make small action changes required to effectively hedge American options. As for a comparison of the quadratic penalty multipliers, it is found that $\kappa = 0.005$ and $\kappa = 0.01$ are optimal, outperforming the BS Delta for both 1 and 3% transaction costs. Agents trained with $\kappa = 0.005$ achieved a higher mean final P&L than those trained with $\kappa = 0.01$, while those trained with $\kappa = 0.01$ achieve a lower standard deviation. It is also worth noting that the training of the agent with a linear multiplier of 0.03 completely diverged, further exploiting the deficiencies of the linear penalty function.

*5.2 Market Calibrated DRL with Weekly Re-Training*

The final objective of this study is to optimize the key result of Pickard *et al.* (2024), wherein market calibrated DRL agents outperform the BS Delta method, by examining the impact of re-training DRL agents with re-calibrated stochastic volatility models each week. Recall that options on 8 symbols, with 5 strikes each are used. Hedging results using the empirical asset path for each respective symbol between the sale date (October 16[th], 2023) and maturity date (November 17[th], 2023) are presented in Table 7. Italicized and underlined are the optimal agent for each case. Note that the average final P&L is averaged across the five strikes for ease of presentation here, but the results for all 40 options are provided in the Appendix. The asset paths for all 8 symbols between October 16[th], 2023, and November 17[th], 2023, are also available in the Appendix.

Table 7: Mean Final P&L across 5 Strikes when Hedging Empirical Asset Paths

| | $\lambda = 1\%$ | | |
|---|---|---|---|
| **Symbol** | **Delta (Weekly Re-Cal)** | **DRL (Weekly Re-train)** | **DRL (Train Once)** |
| AIG | -$0.42 | -$0.17 | *-$0.02* |
| AMZN | -$1.98 | _-$0.78_ | -$1.65 |
| DELL | _-$0.41_ | -$0.46 | -$0.54 |
| GE | -$1.44 | _$0.65_ | -$1.24 |
| PEP | -$0.67 | _-$0.15_ | -$0.62 |
| TSLA | -$6.72 | -$6.85 | _-$4.38_ |
| TXN | -$1.68 | -$1.44 | _-$0.82_ |
| XOM | -$0.60 | _$0.60_ | -$2.12 |
| λ = 3% | | | |
| AIG | -$1.27 | -$0.75 | _-$0.55_ |
| AMZN | -$4.98 | _-$2.49_ | -$4.14 |
| DELL | -$1.13 | _-$1.01_ | -$1.23 |
| GE | -$4.07 | _-$0.57_ | -$4.18 |
| PEP | -$3.26 | _-$1.59_ | -$2.38 |
| TSLA | -$15.46 | _-$12.03_ | -$12.04 |
| TXN | -$5.67 | _-$3.73_ | -$4.60 |
| XOM | -$2.49 | _-$0.15_ | -$1.82 |

The results confirm the hypothesis that DRL agents re-trained each week with re-calibrated models outperform DRL agents only trained on the sale date. Specifically, weekly retrained agents outperform single trained agents in 12 of the 16 cases listed above. Moreover, the DRL agent with weekly re-training outperforms the BS Delta method, which uses the updated volatility from the re-calibrated model, in 14 of 16 cases. This is a very encouraging result, as not only does it reiterate the finding of Pickard *et al.* (2024) that DRL agents may be implemented for hedging real asset paths by training them with market calibrated models, but by re-calibrating the models with higher frequency, DRL performance improves. The use case here for practitioners is evident: given a basket of options and the underlying option data, i.e., the bid-ask spread and the implied volatility surface, DRL agents may be trained each week with the aid of new, recalibrated Chebyshev models to effectively hedge each option in the basket. Moreover, recall that another key of this proposed method is that re-calibrating does not impose longer training times, as each DRL agent is only trained to hedge between successive recalibration points. As such, this work is trivially extended to more frequent recalibration.

## 6 Conclusions

In conclusion, this paper extends the literature pertaining to the hedging of American options with DRL. First, a hyperparameter sensitivity analysis is performed to shed light on hyperparameter impact and provide a guide for practitioners aiming to implement DRL hedging. Results indicate not only the optimal hyperparameter sets for analyses across learning rates and episodes, learning rates and NN architectures, training steps, and transaction cost penalty functions, but provide key insight on what combinations should be avoided. For example, high learning rates should not be used in conjunction with high amounts of training episodes, and low learning rates should not be used when there is few training episodes. Instead, best results stem from middling values, and results also show that if the learning rates are low enough, increasing the number of training episodes will improve results. Further the analysis shows that two hidden layers with 32 nodes is too shallow for the actor and critic networks, and similar to the first experiment it is shown that when the learning rates are lowered, expanding the depth of the NN improves performance. Further, a third experiment shows that caution should be given to using too many training steps to avoid instability. A final hyperparameter test indicates that a quadratic transaction cost penalty function yields better results than a linear version.

This paper then built upon the key results of Pickard *et al.* (2024), by training DRL agents with asset paths from a market calibrated stochastic volatility model, noting that new DRL agents are trained each week to re-calibrated models. Results show that DRL agents trained each week outperform those trained solely on the sale date. Moreover, results show that DRL agents, both the single-train and weekly-train variants, outperform the BS Delta method at 1 and 3% transaction costs. As such, this paper has significant practical relevance, as it is shown that practitioners may use readily available market data to train DRL agents to hedge any option on their portfolio.

**Appendix**

Table A1: Full Results from Hedging True Asset Paths

| Symbol | TC Rate (λ) | Strike | Delta (Weekly Re-Cal) | DRL (Weekly Re-train) | DRL (Train Once) |
|---|---|---|---|---|---|
| AIG | 1 | 50 | $0.04 | -$0.18 | $0.09 |
| | | 55 | -$0.22 | -$0.51 | -$0.06 |
| | | 60 | -$0.88 | -$0.36 | -$0.71 |
| | | 65 | -$0.81 | $0.20 | $0.56 |
| | | 70 | -$0.23 | -$0.02 | -$0.01 |
| | 3 | 50 | -$0.08 | -$0.45 | $0.08 |
| | | 55 | -$0.71 | -$1.17 | -$0.39 |
| | | 60 | -$2.53 | -$1.16 | -$2.34 |
| | | 65 | -$2.79 | -$0.96 | -$0.10 |
| | | 70 | -$0.24 | -$0.02 | -$0.01 |
| AMZN | 1 | 125 | -$2.41 | -$2.28 | -$2.27 |
| | | 130 | -$2.86 | -$1.61 | -$2.75 |
| | | 135 | -$2.04 | -$1.72 | -$2.30 |
| | | 140 | -$1.48 | $1.80 | -$1.37 |
| | | 145 | -$1.11 | -$0.09 | $0.42 |
| | 3 | 125 | -$5.45 | -$4.35 | -$5.07 |
| | | 130 | -$6.30 | -$3.89 | -$5.89 |
| | | 135 | -$5.87 | -$4.19 | -$5.62 |
| | | 140 | -$5.70 | $0.70 | -$4.02 |
| | | 145 | -$1.58 | -$0.73 | -$0.09 |
| DELL | 1 | 60 | -$0.11 | -$0.64 | -$0.11 |
| | | 65 | -$0.69 | -$1.10 | -$1.14 |
| | | 70 | -$0.77 | -$0.66 | -$1.49 |
| | | 75 | -$0.18 | $0.13 | $0.10 |
| | | 80 | -$0.28 | -$0.05 | -$0.05 |
| | 3 | 60 | -$0.58 | -$1.15 | -$0.57 |
| | | 65 | -$1.94 | -$1.96 | -$2.45 |
| | | 70 | -$2.61 | -$1.98 | -$3.12 |
| | | 75 | -$0.24 | $0.11 | $0.03 |
| | | 80 | -$0.29 | -$0.06 | -$0.05 |
| GE | 1 | 100 | -$0.87 | $0.20 | -$0.88 |
| | | 105 | -$1.40 | -$0.11 | -$1.42 |
| | | 110 | -$1.80 | $0.54 | -$1.82 |
| | | 115 | -$2.00 | $1.37 | -$2.01 |
| | | 120 | -$1.11 | $1.24 | -$0.06 |
| | 3 | 100 | -$2.48 | -$1.05 | -$2.48 |
| | | 105 | -$3.92 | -$1.78 | -$3.93 |

|  |  | 110 | -$5.31 | -$1.04 | -$5.35 |
|  |  | 115 | -$6.31 | $0.35 | -$6.30 |
|  |  | 120 | -$2.35 | $0.67 | -$2.84 |
| PEP | 1 | 150 | -$0.21 | $0.00 | -$0.19 |
|  |  | 155 | -$0.43 | -$0.92 | -$0.57 |
|  |  | 160 | -$1.03 | -$0.84 | -$1.44 |
|  |  | 165 | -$0.88 | $0.88 | -$1.07 |
|  |  | 170 | -$0.79 | $0.15 | $0.16 |
|  | 3 | 150 | -$1.05 | -$0.77 | -$0.99 |
|  |  | 155 | -$2.14 | -$2.44 | -$2.44 |
|  |  | 160 | -$4.14 | -$2.79 | -$4.17 |
|  |  | 165 | -$5.09 | -$0.88 | -$4.16 |
|  |  | 170 | -$3.85 | -$1.09 | -$0.14 |
| TSLA | 1 | 210 | -$5.68 | -$5.62 | -$5.13 |
|  |  | 215 | -$6.11 | -$9.03 | -$2.34 |
|  |  | 220 | -$7.80 | -$6.82 | -$2.51 |
|  |  | 225 | -$7.81 | -$10.04 | -$7.68 |
|  |  | 230 | -$6.21 | -$2.74 | -$4.22 |
|  | 3 | 210 | -$13.38 | -$10.22 | -$10.94 |
|  |  | 215 | -$14.15 | -$15.56 | -$8.98 |
|  |  | 220 | -$17.04 | -$11.72 | -$10.66 |
|  |  | 225 | -$17.35 | -$15.39 | -$16.63 |
|  |  | 230 | -$15.39 | -$7.25 | -$12.99 |
| TXN | 1 | 135 | -$0.52 | -$0.05 | -$0.74 |
|  |  | 140 | -$1.08 | -$1.29 | -$0.79 |
|  |  | 145 | -$2.59 | -$1.35 | -$1.68 |
|  |  | 150 | -$2.65 | -$3.50 | -$0.29 |
|  |  | 155 | -$1.55 | -$1.02 | -$0.63 |
|  | 3 | 135 | -$2.51 | -$1.83 | -$3.16 |
|  |  | 140 | -$4.58 | -$3.34 | -$3.50 |
|  |  | 145 | -$8.27 | -$3.31 | -$6.44 |
|  |  | 150 | -$8.61 | -$7.40 | -$4.99 |
|  |  | 155 | -$4.40 | -$2.78 | -$4.92 |
| XOM | 1 | 100 | -$0.16 | $1.12 | -$1.05 |
|  |  | 105 | -$0.13 | $2.44 | -$2.38 |
|  |  | 110 | -$1.15 | -$0.17 | -$5.65 |
|  |  | 115 | -$1.15 | -$0.48 | -$1.24 |
|  |  | 120 | -$0.42 | $0.07 | -$0.29 |
|  | 3 | 100 | -$1.67 | $0.02 | -$0.94 |
|  |  | 105 | -$3.30 | $1.63 | -$1.08 |
|  |  | 110 | -$3.55 | -$0.98 | -$4.15 |
|  |  | 115 | -$2.91 | -$1.36 | -$2.95 |
|  |  | 120 | -$1.00 | -$0.05 | $0.01 |

Table A2: Realized Asset Paths between Sale and Maturity Dates

| Date | AIG | AMZN | DELL | GE | PEP | TSLA | TXN | XOM |
|---|---|---|---|---|---|---|---|---|
| 10/16/2023 | 62.61 | 132.55 | 67.76 | 109.14 | 161.08 | 253.92 | 154.3 | 109.95 |
| 10/17/2023 | 63.18 | 131.47 | 67.9 | 110.02 | 160.37 | 254.85 | 154.26 | 111.39 |
| 10/18/2023 | 62.6 | 128.13 | 66.89 | 107.57 | 162.03 | 242.68 | 151.82 | 112.95 |
| 10/19/2023 | 61.17 | 128.4 | 67.03 | 106.95 | 160.56 | 220.11 | 150.94 | 113.02 |
| 10/20/2023 | 59.52 | 125.17 | 65.91 | 106.08 | 160 | 211.99 | 147.81 | 111.08 |
| 10/23/2023 | 58.98 | 126.56 | 65.54 | 106.69 | 160.08 | 212.08 | 146.32 | 109.45 |
| 10/24/2023 | 59.87 | 128.56 | 66.95 | 113.62 | 162.19 | 216.52 | 146.92 | 108.39 |
| 10/25/2023 | 60.96 | 121.39 | 66.34 | 111.2 | 162.35 | 212.42 | 141.79 | 108.59 |
| 10/26/2023 | 60.85 | 119.57 | 64.51 | 108.8 | 161.41 | 205.76 | 144.01 | 107.6 |
| 10/27/2023 | 59.53 | 127.74 | 65.96 | 106.35 | 159.62 | 207.3 | 143.12 | 105.55 |
| 10/30/2023 | 61 | 132.71 | 66.43 | 109.81 | 162.28 | 197.36 | 140.5 | 105.88 |
| 10/31/2023 | 61.31 | 133.09 | 66.91 | 108.63 | 163.28 | 200.84 | 142.01 | 105.85 |
| 11/1/2023 | 62.05 | 137 | 68.67 | 107.25 | 164.87 | 205.66 | 143.17 | 105.64 |
| 11/2/2023 | 64.36 | 138.07 | 68.65 | 107.78 | 166.83 | 218.51 | 147.31 | 109.11 |
| 11/3/2023 | 63.74 | 138.6 | 69.51 | 108.92 | 166.79 | 219.96 | 150.23 | 107.78 |
| 11/6/2023 | 63.94 | 139.74 | 71.97 | 111.78 | 166.7 | 219.27 | 147.5 | 105.87 |
| 11/7/2023 | 63.1 | 142.71 | 71.61 | 111.8 | 167.18 | 222.18 | 146.59 | 104.21 |
| 11/8/2023 | 62.58 | 142.08 | 72.48 | 113.85 | 167.39 | 222.11 | 145.22 | 102.93 |
| 11/9/2023 | 62.85 | 140.6 | 72.24 | 113.09 | 166.16 | 209.98 | 143.27 | 102.96 |
| 11/10/2023 | 63.82 | 143.56 | 73.5 | 115.27 | 166.92 | 214.65 | 147.19 | 103.75 |
| 11/13/2023 | 63.73 | 142.59 | 73.77 | 115.52 | 167.77 | 223.71 | 145.87 | 104.84 |
| 11/14/2023 | 64.21 | 145.8 | 73.89 | 117.25 | 168.11 | 237.41 | 149.93 | 104.29 |
| 11/15/2023 | 63.48 | 143.2 | 72.94 | 116.3 | 167.25 | 242.84 | 151.87 | 103.66 |
| 11/16/2023 | 64.12 | 142.83 | 72.54 | 118.94 | 167.71 | 233.59 | 151.89 | 102.46 |
| 11/17/2023 | 64.48 | 145.18 | 73.6 | 119.93 | 166.76 | 234.3 | 154.62 | 104.96 |